\documentclass[preprint,review,12pt]{elsarticle}

\usepackage{amssymb}
\usepackage{amsmath}
\usepackage{graphicx}

\journal{Journal of Theoretical Biology}

\begin{document}

\begin{frontmatter}



\title{Stem cell population asymmetry can reduce rate of replicative aging}


\author{Sahand Hormoz}

\address{Kavli Institute for Theoretical Physics, Kohn Hall, University of California, Santa Barbara, CA 93106, USA}

\begin{abstract}
Cycling tissues such as the intestinal epithelium, germ line, and hair follicles, require a constant flux of differentiated cells. These tissues are maintained by a population of stem cells, which generate differentiated progenies and self-renew. Asymmetric division of each stem cell into one stem cell and one differentiated cell can accomplish both tasks. However, in mammalian cycling tissues, some stem cells divide symmetrically into two differentiated cells and are replaced by a neighbor that divides symmetrically into two stem cells. Besides this heterogeneity in fate (population asymmetry), stem cells also exhibit heterogenous proliferation-rates; in the long run, however, all stem cells proliferate at the same average rate (equipotency). We construct and simulate a mathematical model based on these experimental observations. We show that the complex steady-state dynamics of population-asymmetric stem cells reduces the rate of replicative aging of the tissue --potentially lowering the incidence of somatic mutations and genetics diseases such as cancer. Essentially, slow-dividing stem cells proliferate and purge the population of the fast-dividing --older-- cells which had undertaken the majority of the tissue-generation burden. As the number of slow-dividing cells grows, their cycling-rate increases, eventually turning them into fast-dividers, which are themselves replaced by newly emerging slow-dividers. Going beyond current experiments, we propose a mechanism for equipotency that can potentially halve the rate of replicative aging. Our results highlight the importance of a population-level understanding of stem cells, and may explain the prevalence of population asymmetry in a wide variety of cycling tissues.
\end{abstract}


\end{frontmatter}


\section{Introduction}
As cells replicate they accumulate damage, such as somatic mutations, which can eventually manifest as a malady; cancer is the deadliest culmination of such replicative misfortunes \cite{Hanahan00,Bishop95,Vogelstein93,Vogelstein00,Vogelstein04}. Cells that undergo many rounds of divisions are more likely to develop cancer; particularly susceptible are the somatic stem cells of tissues that constantly cycle, i.e. the mammalian intestinal epithelium, epidermis, and germ line \cite{Cairns75,Michor04}.

Stem cells are defined for their capability to generate more stem cells --self-renewal-- and daughter cells that differentiate \cite{Siminovitch63}. Asymmetric divisions of a stem cell into one stem cell and one differentiated cell will satisfy both these objectives: replenish the stem cell pool and generate the differentiated progenies cycling through the tissue \cite{Potten90,Morrison06}. However, lineage-tracing experiments have demonstrated that in a diverse set of mammalian cycling tissues, stem cells divide symmetrically, generating daughter cells that acquire the same fate \cite{Clayton07,Lopez10,Snippert10,Klein10,Simons11}. The balance between proliferation and differentiation is maintained at the population level: some stem cells divide into two stem cells, whereas others generate two differentiated cells and are lost \cite{Klein11}. To maintain homeostasis --constancy in cell number and tissue organization, both outcomes occur with frequency one-half.

The justification for the prevalence of population asymmetry in cycling tissues remains elusive. The assertion that symmetric divisions are required for regenerating the tissue following injury and during development \cite{Morrison06,Watt00}, motivates the capability but not the continual use of this strategy in homeostasis. For instance, although germ line stem cells in the {\it Drosophila} ovaries are capable of symmetric division when a stem cell is lost, during normal proliferation all divisions are asymmetric \cite{Spradling01}. We propose that population asymmetry reduces the rate of replicative aging in mammalian cycling tissue maintained by rapidly dividing stem cells.

Herein, we formulate a model and simulate the complex dynamics of stem cells with population asymmetry and external regulation: stem cells lost to differentiation are replaced by their neighbors. The source of stochasticity in stem cell fate is the asynchronous waiting time before division. We show that heterogenous stem cells --cells that cycle at different rates-- with population asymmetry undergo fewer divisions for the same flux of differentiated cells compared with stem cells that use division asymmetry.

 The degree of heterogeneity does not need to be large: the slowest dividers are only four times slower than the fastest dividers in the population, consistent with experimental observations. A single stem cell's cycling-rate fluctuates, but its long-time average is equal to the population average. All stem cells are therefore equipotent; no single cell has a proliferative advantage over the others in the long run.

Going beyond current experiments, we show that the replicative-aging rate can be further reduced using the following strategy: stem cells by default divide faster than the average rate of cycling in the population. External stimuli slow down the division rate, a trait which is encoded in methylation patterns and passed on to the daughter cells. After a random number of divisions the methylation signal is lost and the daughter stem cells revert to their default behavior. This simple strategy can reduce the rate of replicative aging by a factor of two; we conjecture that this mechanism is used in mammalian cycling tissues.

\section{Results}

We will refer to replicative aging --the cumulative number of divisions undergone by a cell-- simply as `aging'. Cell aging without replication can occur through post-mitotic mutations in mitrochondrial DNA \cite{Wallace99}, stress-induced DNA damage, or accumulation of altered proteins \cite{Kirkwood95,Campisi07}.

If a population of stem cells divides synchronously, all the stem cells will age by the same amount regardless of their ancestry; both division asymmetric and population asymmetric strategies are equivalent. A slight degree of heterogeneity in the division rates, however, can lower the average aging rate, if symmetric divisions are allowed, where some stems cells can replace others (Fig.1). In such a population, the fast-dividing cells take over most of the proliferative burden and are eventually replaced by slow-dividing cells.

\begin{figure}
\begin{center}
      \centerline{\includegraphics[scale=1]{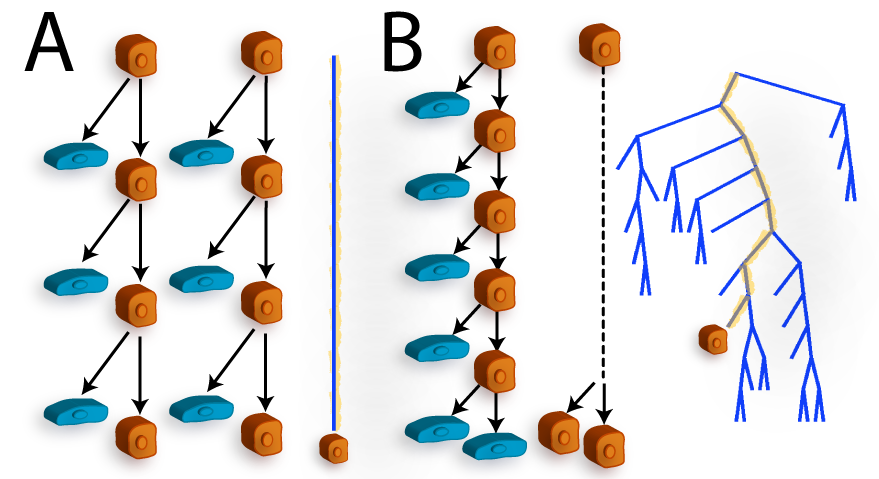}}
       \caption{Heterogeneous proliferation rates. A) Two stem cells (orange) divide asymmetrically three times at a constant rate generating six differentiated progenies (blue). Both stem cells will have aged by the same amount (three divisions). The phylogeny tree of each stem cell --neglecting the branches corresponding to somatic cells-- is simply a straight line. B) The stem cell on the left undergoes four asymmetric divisions followed by a symmetric division into two differentiated cells. Stem cell on the right divides symmetrically into two stem cells at a lower rate. The net product is the same as A. The two stem cells, however, have aged by only one division. With symmetric divisions, the phylogeny of a stem cell is more complicated. An example of a phylogeny tree is shown for stem cells with heterogenous cycling rates and symmetric divisions; terminal nodes correspond to stem cells currently in the population. The highlighted part of the tree depicts a stem cell that has undergone 8 divisions and is younger than some of the other descendants.} \label{Fig1}
\end{center}
\end{figure}

We consider an idealized model of cycling tissues (Methods, Fig.2):  a fixed number of stem cells form an epithelial basal layer \cite{Morrison08}. After a random waiting-time, stem cells divide asynchronously --along their out of plane axis-- into two differentiated progenies, which leave the layer. In response, a randomly-chosen neighbor symmetrically divides in the lateral direction and replaces the lost stem cell --externally regulated division. This model closely resembles the actual dynamics of mammalian cycling tissues \cite{Lopez10,Snippert10,Klein10,Simons11}. Since a division into two differentiated cells is always accompanied by another into two stem cells, the frequency of both types of division is 0.5, ensuring a constant flux of differentiated cells and self-renewal of the stem cell population.

We will address in detail the validity of the assumptions behind this model in the Discussion section. However, we note two key points at the onset. Division into two differentiated progenies can be replaced by a stem cell that simply differentiates and leaves the basal layer. This simpler picture does not change the results. However, as we will show below, it is essential that loss of a stem cell to differentiation trigger symmetric division of a neighboring cell, and not the other way around.

\begin{figure}
\begin{center}
      \centerline{\includegraphics[scale=1]{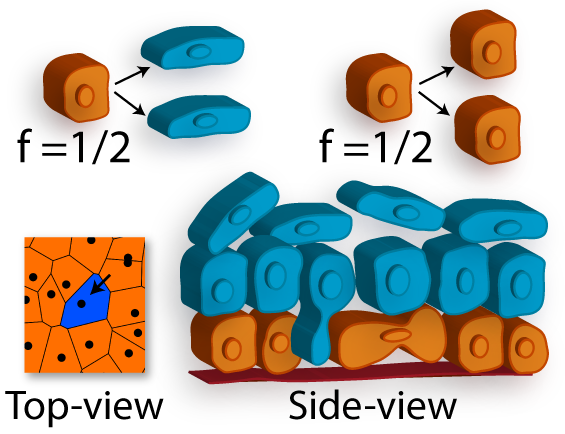}}
       \caption{Population asymmetry. Stem cells form an epithelium attached to the basal membrane. A stem cell divides symmetrically into two differentiated cells and leaves the basal layer. The lost stem cell is replaced by one of its neighbors which divides laterally into two stem cells. All division are symmetric; each type occurs with frequency $1/2$. The net product of the two divisions is two differentiated cells and two stem cells. Above, the third stem cell from the left has divided symmetrically into two differentiated cells and is leaving the basal layer. Its neighbor to the right is dividing symmetrically into two stem cells to replace the lost cell.} \label{Fig2}
\end{center}
\end{figure}

The above proliferation strategy results in dynamics that closely resemble the `stepping-stone' model of neutral drift in population dynamics \cite{Kimura64,Klein11} (see also `voter' models in statistical physics \cite{Holley76}). If we initially label all cells in the population with a different color (Fig.3), overtime, we observe that most cells leave the stem cell layer, and are replaced by their neighbor. Those that remain, grow in number, forming clusters of daughter cells (of the same color) called clones. If all stem cells are equivalent and cycling at the same rate, no clone is more likely to survive than its neighbors. The boundaries between adjacent clones diffuse back and forth, roughly expanding as $\sqrt{t}$; the size of the surviving clones therefore increases as $t$ (this argument only holds in 1 and 2 dimensions, for a rigorous derivation and the log correction to this scaling in 2D, see \cite{Bramson80}). Eventually, in a finite population, one clone will take over, achieving fixation; in Fig.3, a stem cell initially labeled red fixates. The resulting distribution of clone sizes is self-similar: the distribution of the ratio of clone sizes to the average clone size at any given time is constant. Statistical signatures of clone sizes have been used to experimentally confirm the above dynamics in cycling tissues in vivo \cite{Klein11}.

\begin{figure}
\begin{center}
      \centerline{\includegraphics[scale=1]{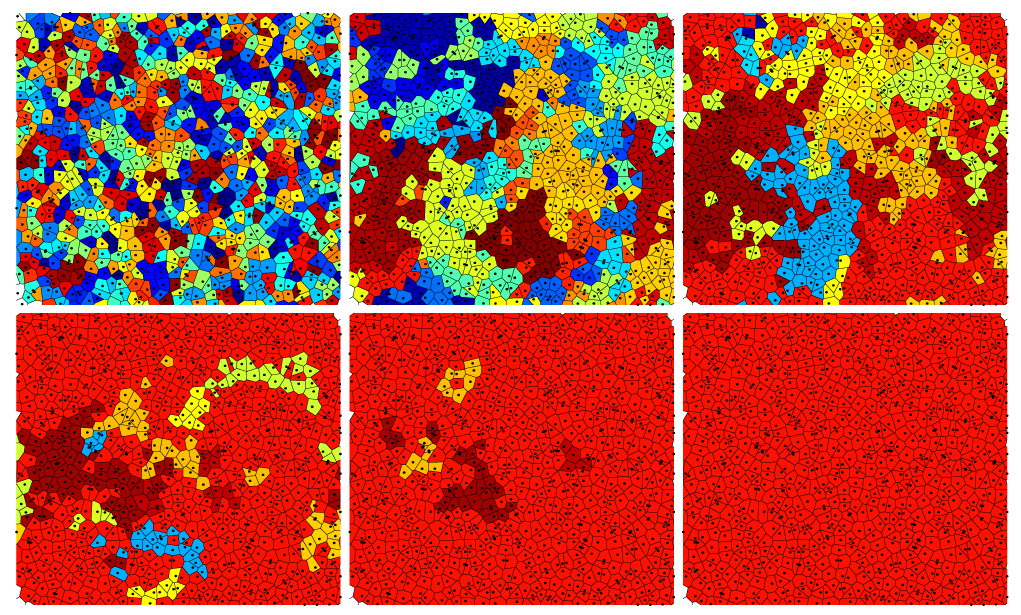}}
       \caption{Clonal expansion and loss of heterogeneity. Cells are initially colored at random and their evolution followed. Color of the daughter cells is inherited from the parent. Eventually the clone of a red stem cell fixates. The time steps from left to right and top to bottom are, t=1, 20, 60, 100,140, 200, and 260. The stem cells have heterogenous division rates, $\sigma = 0.15$.} \label{Fig3}
\end{center}
\end{figure}

\subsection{Heterogeneity} 
Next, consider a heterogenous population, where stem cells cycle at different rates: stem cell $i$ terminally divides into differentiated progenies at rate $r_i$ drawn from a Gaussian distribution with mean 0.5 and standard deviation $\sigma$. The cycling-rate is inherited by the daughter cells following a division into two stem cells. The average cycling-rate of the population is held fixed to ensure a constant flux of one differentiated cell per stem cell per unit of time (Methods). 

We emphasize that cycling refers to a stem cell dividing into differentiated progenies and its subsequent replacement by a neighbor. The term slow-dividers will refer to cells that are less likely to divide into differentiated progenies and leave the stem cell population in a given interval of time. As the name implies, a cluster of slow-dividers will also undergo fewer stem cell to stem cell divisions in a give period of time since fewer cells need to be replaced. However, stem cells will always divide to replace a lost neighbor; the rate of this division is not related to the cell's own cycling rate, $r$, but indirectly to that of the neighbor. This distinction is important when a slow-divider is adjacent to a fast-divider. Alternatively, we can think of $r$ as the rate of differentiation, which indirectly sets the division rate of a cell through its neighbors. 

The clones are no longer equivalent; some are descendants of fast-dividing cells and others of slow-dividing cells. Clusters of fast-dividing cells contribute more to the flux since they cycle more quickly. However, at their interfaces, the slow-dividing clones invade the fast-dividing clones, since at any time-step, the fast-dividing cells are more likely to undergo terminal differentiation and be replaced by a neighbor. The fast-dividing cells that are contributing more to the flux --and thereby aging faster-- are systematically removed from the population and replaced by younger cells. 

Define the population-averaged aging rate $\rho(t)$ as the ratio of the total number of divisions starting from time $t=0$ undergone by the stem cells present in the population at time $t$ to the total number of differentiated cells produced since $t=0$. $\rho=1$ for homogenous population --where all cells proliferate at the same rate, since all the stem cells contribute to the flux equally and on average one differentiated cell is produced by each division. Similarly, $\rho=1$ for a population with division asymmetry. Surprisingly, simulations show that the aging-rate with heterogeneity is lower than that of a homogenous or division asymmetric population, $\rho(t)<1$, at early times (Fig.4A); eventually, $\lim_{t \to \infty} \rho \to 1$.

We can understand the eventual reversion to $\rho=1$ by considering the dynamics more carefully. The initially heterogenous population of cells first gives rise to an abundance of slow-dividing cell, since they are more likely to invade their neighbors. As the number of slow-dividing cells grows --clonal expansion, their cycling-rate increases to ensure a constant flux of differentiated progenies. The initially Gaussian distribution of cycling-rates in the population develops a spike at the slowly-dividing rates, the spike grows and reverts to the mean (Fig.4B). Eventually the distribution becomes a $\delta$-function at 0.5 when fixation has occurred and every stem cell has the same proliferation rate.

\begin{figure}
\begin{center}
      \centerline{\includegraphics[scale=1]{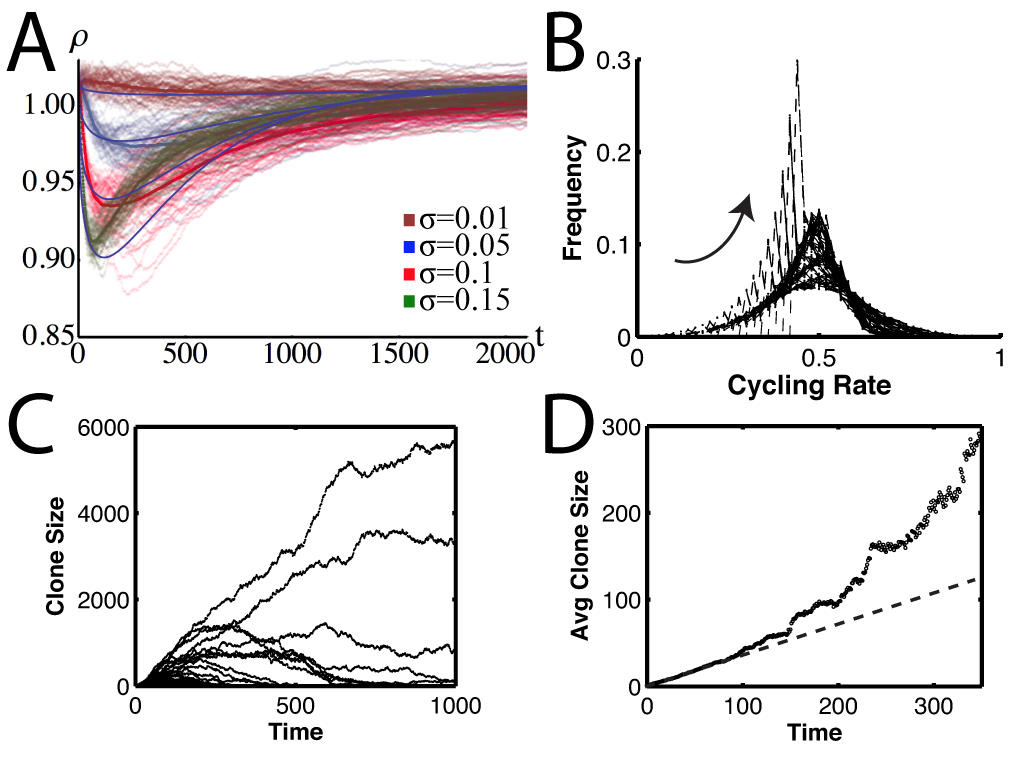}}
       \caption{Heterogeneous but not equipotent. A) Population-averaged aging rate $\rho$ as a function of time. Initially $\rho<1$ before reverting to its asymptotic value of 1. 50 runs are superimposed for each value of $\sigma$. The thick colored lines are the average over the runs. Solid-blue lines are the analytical results of the phenomenological model. B) Frequency of observing a cell with cycling-rate r for the first 120 time steps superimposed ($\sigma = 0.15$). The distribution narrows over time, eventually becoming a $\delta$-function as one clone fixates. C) Clone sizes in the population as a function of time for the same parameter values. Some clones expands as others shrink. D) Average clone size as a function of time. There is a significant deviation at early times from the linear prediction of the neutral drift model.} \label{Fig4}
\end{center}
\end{figure}

\subsection{Phenomenological model}

A simple phenomenological model can elucidate the essential mechanisms for lowering the rate of aging. We include the following key facts in the model: 1) The boundaries of the slow-dividing clones grow faster than would be expected from neutral dynamics (Fig.4D). At early times, the boundaries grow on average with velocity $0.5 - r$, where $r$ is the cycling-rate. 2) As the number of slow-dividing cells increases, their cycling-rate also increases, approaching the population-average rate, 0.5. We assume that the fractional rate of increase of the cycling-rate $r$ is the same as the fractional increase in the number of cells dividing at rate $r$.

Using a Lagrangian framework, we follow the evolution of the cells that at $t=0$ cycled at rate $r_0$; at time $t$, the number of these cells in the population is denoted as $N(r_0,t)$ and their cycling-rate as $r(r_0,t)$. In 2D, the dynamical evolution is given by,

\begin{equation}
\frac{1}{2\sqrt{N(r_0,t)}}\frac{\partial N(r_0,t)}{\partial t} = \frac{1}{2} - r(r_0,t) \label{a1}
\end{equation}
\begin{equation}
\frac{\partial r(r_0,t)}{\partial t} = \frac{1}{N(r_0,t)}\frac{\partial N(r_0,t)}{\partial t}r(r_0,t) \label{a2}
\end{equation}

The number of stem cells cycling at rate $r'$ at time $t$ is
\begin{equation}
\mathcal{N}(r',t) = \int N(r_0,t)\frac{1}{|\frac{\partial r_0}{\partial r}|_t} \delta(r'-r(r_0,t)) dr_0.
\end{equation}

Eqs. \ref{a1} and \ref{a2} can be combined into the following non-linear ODE --denoting $N(r_0,t)$ as $N$ for simplicity,
\begin{equation}
\frac{d^2 N}{dt^2} = \bigg(\frac{1}{2N} - \frac{1}{\sqrt{N}}\bigg)\frac{dN}{dt} + \frac{1}{N}\bigg(\frac{dN}{dt}\bigg)^2. \label{eqODE}
\end{equation}

The initial distribution of proliferation rates in the simulations is a Gaussian with mean 0.5 and standard deviation $\sigma$. For simplicity, we follow two population of cells from this distribution: fast-dividing cells cycling at rate $r_f = 0.5+\sigma$, and slow-dividing cells cycling at rate $r_s = 0.5-\sigma$. Simulations (Fig.4) show a monotonous decrease in the second moment of the cycling-rate distribution due to loss of heterogeneity --the mean is fixed by construct; the third moment, however, peaks when $\rho$ is at its minimum. We claim that to the lowest order, the rate of aging at time $t$ is a function of the third moment of the distribution of the cycling-rates in the population at time $t$. The more lop-sided the distribution, the more prevalent the presence of slow-dividing clusters that invade their fast-dividing neighbors, ensuring $\rho<1$. The analytical estimate of $\rho(t)$ is

\begin{equation}
1 - \rho(t) \sim \big[ N(r_s,t)\big( \frac{1}{2} - r(r_s,t) \big)^3 -N(r_f,t)\big( r(r_f,t) - \frac{1}{2} \big)^3  \big]^{\frac{1}{3}} \label{rho}.
\end{equation}

The exact shape of the curve $\rho(t)$ depends on the system size. Loss of heterogeneity --rate of reversion of $\rho(t)$ to its asymptotic value-- will be slower for larger systems since there is a larger pool of initially slow-dividing cells. The analytical theory is a good fit to the simulations results (Fig.4A, Fig.7C); only two fitting parameters are used (Methods).

\subsection{Heterogeneous and equipotent}

To sustain the low rate of aging, we need to ensure that heterogeneity is not lost by clonal expansion, and somehow reintroduced into the system. Inspired by methylation as a method of passing epigenetic information to daughter cells (see Discussion), we assume that after a random number of divisions the inherited methylation patterns are lost and the cell's cycling-rate randomized (see Methods for implementation). The average number of divisions (time-scale) to losing inherited information is denoted as $\tau_{eq}$. If we average the behavior of the cells over time-scales much longer than $\tau_{eq}$, all cells are equipotent: there are no slow or fast dividers, all cells proliferate at the average rate. With equipotency, $\rho$ reaches a steady-state value $\rho_{\infty}$, which is approximately equal to $\rho(\tau_{eq})$ in the heterogenous but non-equipotent case (Fig.5A).

A steady-state rate of aging $\rho_{\infty}<1$ does not imply that there is a steady-state distribution of cycling-rates, i.e. a constant number of slow and fast dividers. The dynamics is more complicated: as in the non-equipotent case, slow dividers expand as their cycling-rate increases (Fig.5B). Equipotency, however, ensures that somewhere else in the population, a new slow-divider emerges and expands, continuing the cycle (Fig.6). Time-averaging the cycling-rate distribution for a sufficiently-long periods, yields a static skewed distribution --non-zero third moment (Fig.5C). There is an abundance of slow-dividers (tail of the distribution) which are compensated by an abundance of cells that divide only slightly faster than the average.

Equipotency also restores the self-similar dynamics; stem cells behave like a homogenous population for time scales larger than $\tau_{eq}$. For example, the self-similar character of the clone size distributions, and the linear growth of the average clone size is restored (Fig. 5D), consistent with experimental observations \cite{Clayton07,Lopez10,Snippert10,Klein10,Simons11,Klein11}.

\begin{figure}
\begin{center}
      \centerline{\includegraphics[scale=1]{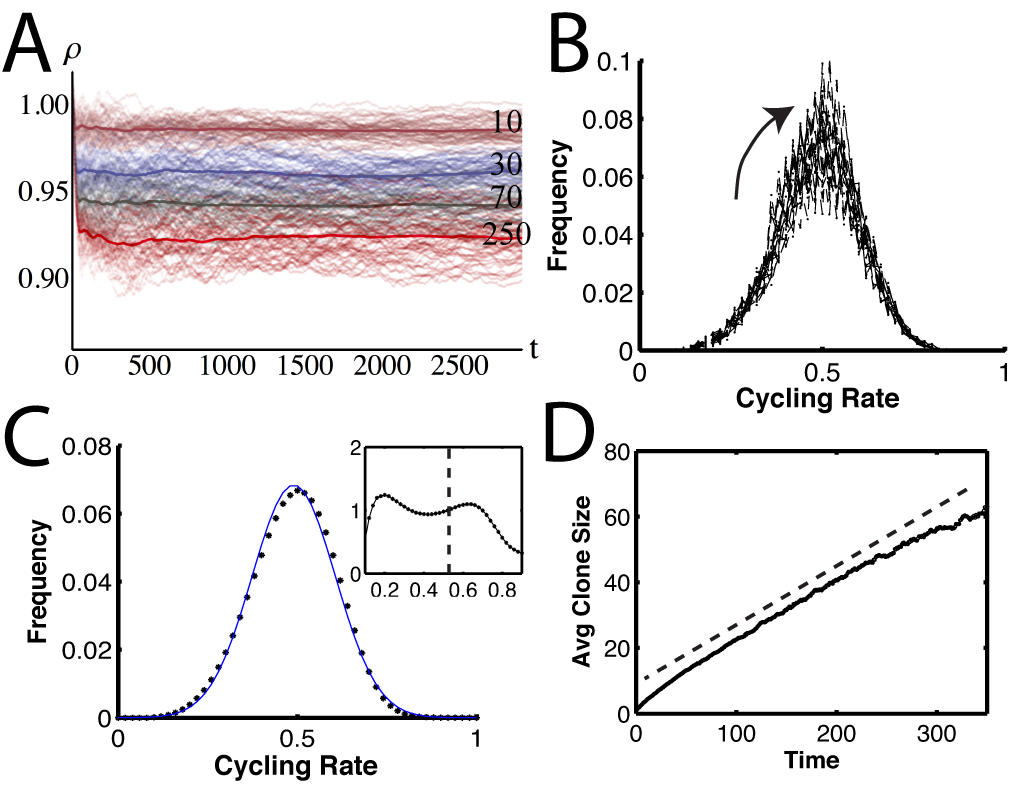}}
       \caption{Heterogenous and equipotent. A) Aging rate $\rho$ as a function of time for $\sigma = 0.15$ and four different values of $\tau_{eq}$; 50 runs and their average. B) Distribution of cycling-rates superimposed for 20 time steps (3981 to 4000) for $\tau_{eq} = 250$. At steady-state, waves of slow dividers are generated and sent towards the mean, where they annihilate (see Fig.6). C) Distribution of proliferation-rates averaged for 1000 steps (3001 to 4000). The time-averaged distribution looks Gaussian but is asymmetric (non-zero third moment). The inset shows the ratio of the actual distribution to a Gaussian fit. There is an excess of slow-dividers --at the tail on the left-side. D) Average clone size as a function of time for $\tau_{eq} = 1$. The self-similar behavior of the neutral drift models is recovered.} \label{Fig5}
\end{center}
\end{figure}

\begin{figure}
\begin{center}
      \centerline{\includegraphics[scale=1]{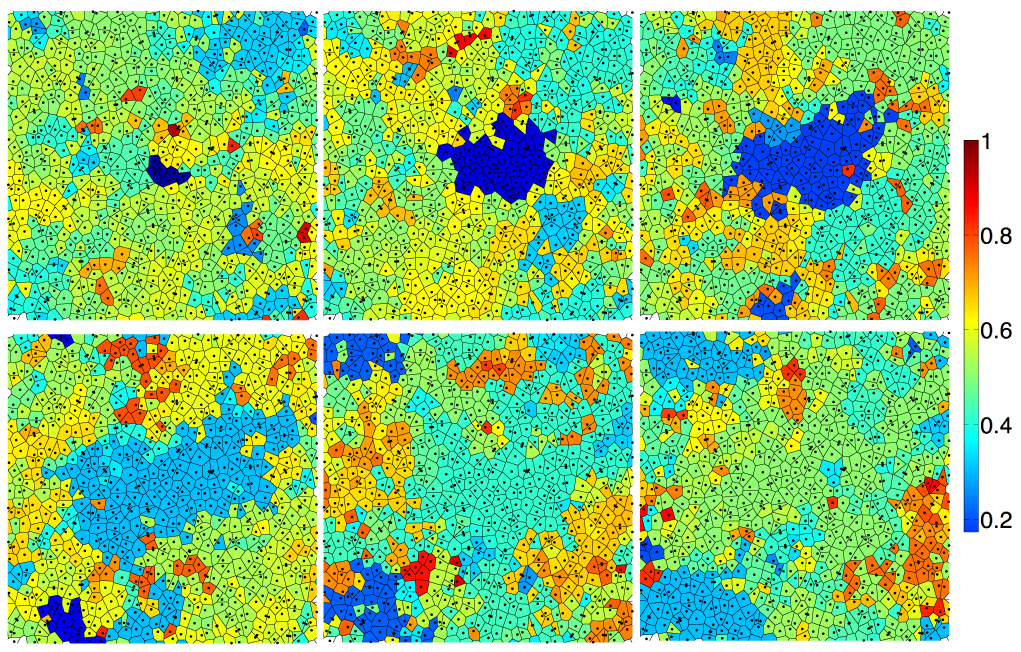}}
       \caption{Maintaining heterogeneity. The color-coding corresponds to the cycling-rate of each cell. Each cell randomizes its cycling rate (drawn from a normal distribution with $\sigma = 0.15$) for $\tau_{eq}  = 50$. A slow-divider in the middle proliferates while increasing its cycling-rate, eventually reaching the mean rate 0.5. A new cluster of slow-dividing cells has formed in the bottom-left corner and is expanding. The time steps shown are from left to right and top to bottom, t=90 to 140 in increments of 10.} \label{Fig6}
\end{center}
\end{figure}

\subsection{Optimizing}
If cycling-rates randomize too quickly, slow-dividing cells do not have sufficient time to replace their faster dividing neighbors. If the randomization rate is too slow, the slow-dividing clones expand too much, reducing the heterogeneity, and increasing the rate of aging. There is an intermediate optimal time-scale for equipotency. We examined the dependence of the steady-state value of aging rate $\rho_\infty$ on $\tau_{eq}$ (Fig.7A). The optimal value of $\tau_{eq}^*$ that minimizes aging rate is approximately the time required to achieve the transient minimum $\rho$ in a heterogenous non-equipotent population; $\tau_{eq}^*$ is the point of extremum in Eq.\ref{rho}: $\frac{d\rho (t)}{dt}|_{\tau_{eq}^* \approx 0}$.

To further optimize $\rho$, we consider other feasible mechanisms:

{\bf Bimodal distribution of cycling-rates}. Thus far, equipotency was implemented by randomizing cycling-rates after a few divisions. Can a different mechanism do better? In Fig.7B, cells establish equipotency by resetting their cycling-rate to a fixed value above-average $\bar{\sigma}+0.5$. Physically, this corresponds to stem cells that are programmed by default as fast-dividers. However, global signaling establishes a methylation pattern where most cells divide slower; after a random number of divisions the cell loses the methylation program and reverts to its default behavior.

This approach results in a bimodal distribution of cycling-rates: slow-dividing cells that replace their neighbors and proliferate, balanced by the fast-dividing cells that are constantly injected into the population when cells reset their cycling-rates. For optimal equipotency time-scale, $\tau_{eq}^*$, and a factor of 6 difference between the proliferation-rates of the fastest and slowest dividers, we find $\rho_{\infty} = 0.4$. This, implies approximately a two-order of magnitude (99\%) reduction in incidence of five somatic mutations --on average required for onset of cancer-- compared with homogenous or division-asymmetric populations (Methods). For our choice of parameters, the optimal time-scale for equipotency is now much shorter (around 3-5 divisions). Quick addition of fast-dividers is required to balance the growing number of slow-dividing stem cells. This strategy is so effective that we conjecture that it is potentially implemented in cycling tissues (see Discussion).

{\bf Biassed replacement}. We consider the possibility that beside terminal differentiation a stem cell can also proliferate by forcing the terminal differentiation of a neighbor. Fig.7C shows the aging-rate where a stem cell cycling at rate $r$, proliferates with probability $1-r$. A modest improvement is observed (4\% for our choice of parameters) since the fast-dividing cells are even more likely to be purged from the population.

{\bf One-dimension}. Finally, we consider a population of stem cells constrained to one-dimension. Although the steady-state aging rate $\rho_{\infty}$ is not significantly different from that of the two-dimensional case, the dynamics of clonal expansion and loss of heterogeneity is qualitatively different (Fig.7D). For little or no heterogeneity, the clone boundaries expand by diffusion: the average surviving clone size increases as $\sqrt{t}$ (as opposed to $t$ in 2D). As heterogeneity increases the boundaries drift rather than diffuse and the clone size scaling becomes closer to linear. The slower clonal expansion implies that heterogeneity is lost at lower a rate, which maximizes the transient period where $\rho<1$. 1D systems can help optimize the aging rate for finite durations for systems without equipotency. 

\begin{figure*}
\begin{center}
      \centerline{\includegraphics[scale=1]{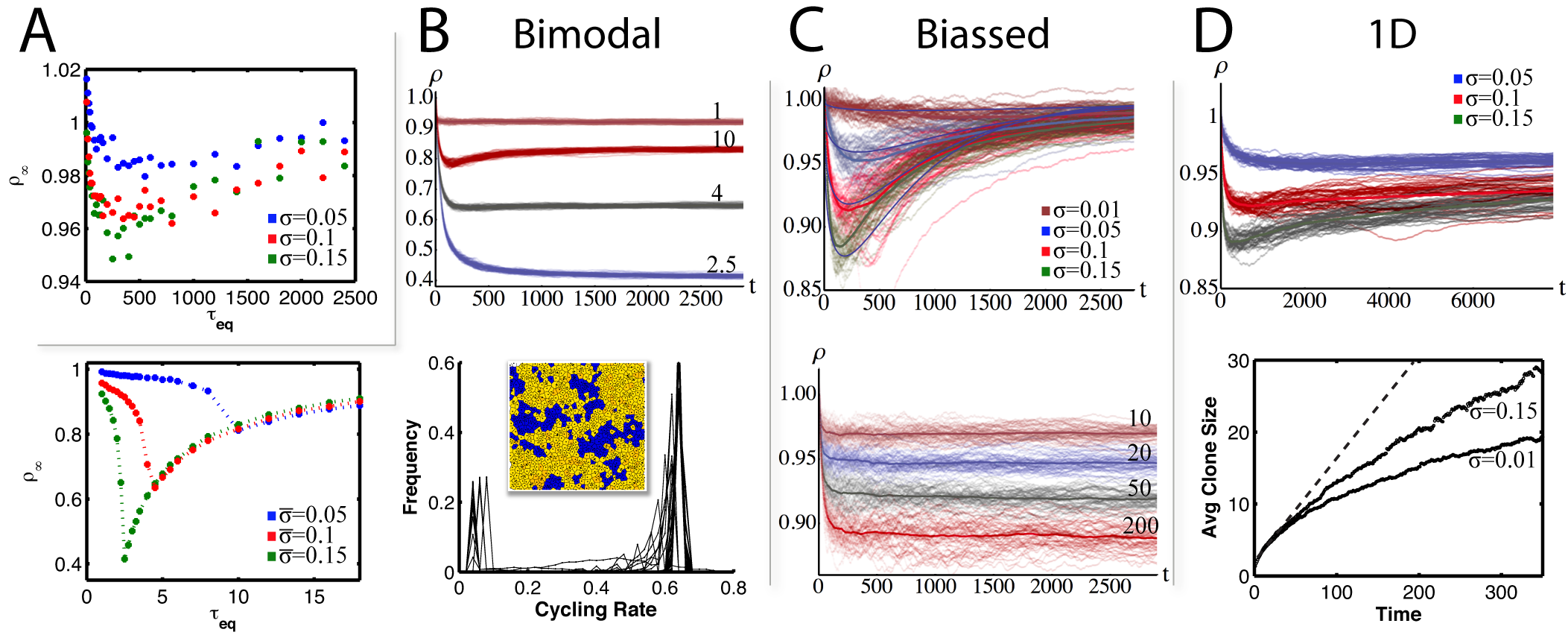}}
       \caption{Minimizing rate of aging. A) Asymptotic value of aging-rate $\rho$ as a function of $\tau_{eq}$ for different degrees of heterogeneity $\sigma$. Optimal $\tau_{eq}$ is at some intermediate value. B) Instead of randomizing, the cell's cycling-rate is reset to $\bar{\sigma}+0.5$. (Top) $\rho$ as a function of t for different values of $\tau_{eq}$; $\bar{\sigma} = 0.15$. Minimum $\rho$ occurs at the high equipotency rate of $\tau_{eq} = 2.5$. The bimodal distribution of the division rates is shown on the bottom --every 5th step for the first 2000 iterations: cells are either fast dividers or slow dividers. The inset shows the spatial heterogeneity of the two types of stem cells (color-coding as in Fig.6). On the bottom-left, we plot $\rho_{\infty}$ as a function of $\tau_{eq}$ for various value of $\bar{\sigma}$. C) Decision of fate is biassed so that slow-dividing cells are more likely to replace a neighbor. (Top) $\rho$ as a function of time and no equipotency. The solid blue lines are the fits from the phenomenological model (Methods). (Bottom) Equipotency for four different values of $\tau_{eq}$; $\sigma = 0.15$. D) 1D system of size 10000x1. (Top) $\rho$ as a function of time. Loss of heterogeneity (clonal expansion) is slower in 1D. (Bottom) Average clone size as a function of time for two values of $\sigma$. The clones grow sub-linearly. B) Top, C), and D) Top, multiple trajectories are 50 runs and their average (thick line).} \label{Fig7}
\end{center}
\end{figure*}

\section{Discussion}
We found that population asymmetry in stem cells with heterogeneous division rates and external regulation can substantially reduce the rate of replicative aging; potentially lowering incidence of somatic mutations and proliferative-maladies such as cancer. Our analysis revealed a non-trivial steady-state dynamics: slow-dividing cells constantly replace their fast-dividing (older) neighbors who had shouldered the brunt of the proliferative demand. A key requirement was equipotent stem cells: in the long run no cell had a proliferative advantage over the others; every cell divided at the same average rate. Equipotency was enforced by assuming that after a random number of divisions, the resulting daughter cells lost their inherited traits. Going beyond current experiments, we conjectured that equipotency resets the cycling-rate of the cells to a default value that is larger than the mean division-rate of the population. 

Stem cell population asymmetry using external regulation --where terminal differentiation of a stem cell triggers the expansion of a neighbor-- has been verified using lineage-tracing experiments in the mammalian intestinal crypt and germ line \cite{Lopez10,Snippert10,Klein10,Simons11}. These quantitative studies have demonstrated that these tissues are maintained by an equipotent quickly-cycling population of stem cells. Do these populations also exhibit heterogeneity? Direct observation of heterogeneity in the intestinal epithelium was made by Pruitt et al. \cite{Pruitt10} who using tamoxifen induced Cre-recombinase activity on the Mcm2 gene observed cell-cycle times ranging from 1 to 4 days, consistent with the degree of heterogeneity assumed above. An even larger discrepancy in cycling rates is exhibited by progenies of the $Lgr5^{+}$ stem cells in the intestinal crypt that remain dormant (for weeks) as precursors to the secretory cells, but are capable of reverting back to a cycling stem cell \cite{Buczacki13}.

Despite a lack of direct evidence of heterogeneity of cycling-times in the germ line, there are observations of heterogeneous gene-expression in undifferentiated cells of the mammalian spermatogonia \cite{Nakawaga10}. Two types of undifferentiated cells $A_s$ and $A_{pr}$ show an expression level low in $Ngn3$ and high in $GFR\alpha 1$, whereas undifferentiated $A_{al}$ cells have the opposite expression levels. $A_{al}$ cells are thought to be primed for differentiation and could correspond to the fast-cycling cells in our model. Short-term lineage tracing experiments have shown that the expression patterns can reversibly change with $GFR \alpha 1^{lo}$/$Ngn3^{hi}$ cells becoming $GFR \alpha 1^{hi}$/$Ngn3^{lo}$, ensuring equipotency \cite{Simons11}. When $A_s$ and $A_{pr}$ cells commit to differentiation, they are replaced by fragmentation of a neighboring $A_{al}$ cell that reverses its expression patterns \cite{Klein10}. However, the inter-conversion rate between $A_{al}$ cells and the undifferentiated cells $A_s$ and $A_{pr}$ is comparable to their cycling rates \cite{Nakawaga10}; the equipotency time-scale $\tau_{eq}$ might be too short for a noticeable impact on the rate of replicative aging.

Intestinal crypt stem cells are also believed to exhibit dynamics where the terminal differentiation and loss of one stem cell results in expansion of a neighbor. Wnt signaling \cite{Flier09,Gregorieff05} --in particular a global Wnt gradient along the crypt \cite{Batlle02}-- is thought to control the average proliferation rate \cite{Buske11}; as do insulin-like growth factors like IGF1 \cite{Pruitt10}. Nearest-neighbor communication seems to be mediated by Notch signaling and lateral inhibition. In the crypt, stem cells laterally inhibit their neighbors to ensure that the neighbor's fate at division is differentiation \cite{Buske05}. A biological mechanisms for the biassed expansion proposed above (Fig.7C) might be the build-up of excess Notch ligands by slow-dividing cells. 

We assumed that cycling-rates were passed on to daughter cells after division. Moreover, to establish equipotency, we required that the rates randomize after a few divisions; the inherited information forgotten. Both of these requirements can be implemented through epigenetic methylation tags \cite{Pfeifer90,Silva93}. Yatabe et al. \cite{Yatabe01} studied methylation patterns in the human colon crypts and observed somatic inheritance, random methylation changes, and evidence of population asymmetry.

Our model is an idealized picture of stem cell proliferation. We only focused on aging in the somatic stem cells and not the differentiated progenies as they are continually discarded. We also assumed that stem cells divided symmetrically into two differentiated cells. The product of this division can also be thought of as transit-amplifying (TA) progenies that undergo a few more rounds of division before terminal differentiation. Including TA cells as the differentiation product of rapidly cycling stem cells does not change the relative improvement observed using population asymmetry over division asymmetry. 
Of course, a hierarchical strategy, where extremely slowly-dividing (quiescent) stem cells give rise to TA cells that undergo many rounds of division before terminal differentiation \cite{Cairns75}, significantly reduces the aging rate of the stem cells --regardless of the division strategy. However, experimental evidence  \cite{Clayton07,Lopez10,Snippert10,Klein10,Simons11} suggests that many mammalian cycling tissues are supported by a population of rapidly dividing stem cells.

We assumed that after a random waiting time a stem cell divides into two differentiated cells. However, it is not clear that differentiation is tied to cell cycle. Our results will not change if stem cells simply differentiate and leave as opposed to dividing into differentiated cells that leave. Heterogeneity in cycling-rate will then be a consequence of heterogeneity in dwell-time before differentiation. However, it is essential that differentiation triggers the proliferation of a neighbor. If, on the contrary, differentiation is triggered by a neighbor's division into two stem cells, the opposite behavior emerges --aging rate increases. We also assumed that all divisions were symmetric; stem cell to two stem cells, or stem cell to two differentiated progenies. A mixed strategy can also be considered where some fraction of divisions are asymmetric. This will not change our results qualitatively. However, the observed improvement in aging rate decreases as the fraction of asymmetric divisions increases.

The geometry of a 2D and 1D epithelium of stem cells used in the model is an over-simplification of real tissues. Although stem cells seem to be predominantly confined to a thin --few cells deep-- layer in cycling tissues, they are usually interspersed with other types of somatic cells. For example, stem cells at the base of the crypt (Lgr5 cells) are juxtaposed by Paneth cells, and in the germ line by large somatic Sertoli cells \cite{Simons11}. A more comprehensive model should account for this added complexity, however, the results will not be qualitatively different. Previous models of population asymmetry that have relied on effective simplified `lattices' of stem cells are in good agreement with experiments \cite{Klein11}.

Notion of heterogeneity can be employed in a different context. Most populations of stem cells contain quiescent cells \cite{Fuchs09,Li10} that cycle so slowly that are effectively dormant. Despite suggestions that quiescent stem cells can increase longevity by replacing actively-dividing stem cells \cite{Li10}, lineage-tracing experiments \cite{Simons11,Klein11} suggest that these cells play little to no role in homeostatic cycling of tissues. Rather, they seem to be prominent in tissue repair following a catastrophic perturbation  \cite{Kelley2012} and during development \cite{Fuchs09}. The heterogeneity used in our model is subtle and the slow-dividing cells are in no way quiescent. To ensure that extremely slow-dividing cells were not essential for our results, we repeated the simulations while constraining the cycling-rates to $\pm 3 \sigma$ and observed the same reduction in the aging rate. For instance, constraining the division probability of the bimodal strategy (Fig. 7B) to 0.2-0.8, resulted in a population where the fastest dividers proliferated only 3 times faster than the slowest dividers; the dynamics still generated $\rho = 0.87$, a potential 50\% reduction in incidence of cancer.

Our results motivate a population level understanding of stem cells. Because of heterogeneity and complex dynamics, static gene-expression profiles provide limited insight into stem cell phenotypes. Study of population epigenetics and population dynamics of stem cells can bring fresh insight into the delicate interplay that occurs in homeostatic cycling tissues.

\section{Methods}

\subsection{Simulations} 
An asynchronous cellular automata algorithm was used: A 2D hexagonal lattice of $N$ points with periodic boundaries is generated and perturbed by adding a random number drawn from a Gaussian distribution with zero mean and standard deviation 30\% of the lattice constant to the x and y coordinates of each point. The $N$ cells are constructed by a Voronoi decomposition of the randomized lattice; two cells that share a common edge are denoted as neighbors. Each cell is initially assigned a cycling-rate $r$ --corresponding to a Poisson rate parameter-- drawn from a Gaussian distribution with mean 0.5 and standard deviation $\sigma$.

In each time step, the following sequence is executed $N$ times: 
1) A cell is chosen at random from the lattice. With probability $r$, it divides terminally into two differentiated cells.
2) If the division occurs, one of its neighbors is chosen at random to divide into two stem cells and replace the lost cell. The daughter cells inherit the cycling rate, and the age (number of prior divisions plus one) of their parent.
3) If equipotency is turned on, at every division with probability $N_e/\tau_{eq}$ --where $\tau_{eq}$ is the time-scale for equipotency and $N_e$ the number of steps required prior to equipotency-- the cell advances one step to resetting its cycling rate. After completing $N_e$ steps, the cycling rate of the cell is redrawn from a Gaussian distribution with mean 0.5 and standard deviation $\sigma$. The new cycling rate is inherited by all the daughter cells in subsequent divisions. At the end of each time step, every cell's cycling rate is modified by the same amount to ensure that the population-average  of $r$ is 0.5. On average, at every time step a total of $N$ differentiated cells are generated.

Above algorithm was modified for two cases: 1) In Fig. 7B (Optimization) equipotency is introduced by always resetting the cell's cycling rate to $0.5+\bar{\sigma}$. Moreover, the cycling rates are constrained to the range $\pm 3 \bar{\sigma}$; 2) In Fig. 7C (Optimization) the selected cell divides symmetrically into differentiated cells (step 1 above) with probability $r$; with probability $1-r$ it divides into two stem cells and removes a random neighbor. 3) In Fig. 7D (Optimization) the above algorithm was modified to run for a one-dimensional lattice. $N_e = 1$ for all plots with equipotency, except for Fig. 7A, where $N_e = 10$. The algorithm was implemented in Matlab R2011b. We used a 100x100 and 10000x1 lattice for the 2D and 1D simulations respectively (N = 10000). For generating the mosaic plots of the lattices (Fig. 3, 5, inset in 7B) a 30x30 lattice (N=900) was used.

\subsection{Estimating incidence of cancer} Cells need to accumulate a series of independent mutations for tumorigenesis \cite{Vogelstein93}. Assume each step (mutation) $i$ is an independent Poisson process with rate $k_i$. The probability that step $i$ will occur in the time interval $t$ to $t+dt$ is an exponential distribution, $w_i(t) = k_i e^{-k_it}$ --known as dwell time. The probability that step 2 occurs at time $t$ is contingent on step 1 having occurred at some time before $t'<t$, $p_2(t) = \int_0^t w_1(t') w_2(t-t')dt' = (w_1 \otimes w_2)(t)$, where $\otimes$ is the convolution operator. Similarly, probability of completing $N$ steps at time $t$ is given by, 
\begin{equation}
p_N(t) = (w_1 \otimes \hdots w_i \otimes \hdots w_N)(t).
\end{equation}
Assuming that cancer is an unlikely event ($k_i t \ll 1$), we can use a Taylor expansion in $t$ (Tom Chou, private communication),
\begin{equation}
p_N(t) \approx \frac{t^N}{(N-1)!} \prod_{i=1}^N k_i.
\end{equation}
$p_N(t)$ is the incidence of cancer; the probability that an individual of age $t$ develops cancer in the time interval $dt$. Heterogeneous population-asymmetric stem cells will age at a lower rate, $k'_i = \rho k_i$. The likelihood of incidence of cancer is smaller by a factor of $\rho^N$. For $\rho$= 0.9, and 0.4, and a typical value of $N=5$ \cite{Vogelstein93}, incidence of cancer is roughly reduced  by 40\%, and 99\% respectively.

\subsection{Fitting the data}
To fit the simulation, we numerically solved Eq.\ref{eqODE} using the initial condition $N(r_{s/f},t=0) = N G(0.5 \pm \sigma)$, where $G$ is a Gaussian distribution with mean 0.5 and standard deviation $\sigma$, and $N$ is a system-size dependent fitting parameter. The other fitting parameter is the proportionality constant $\alpha$ in Eq.\ref{rho}. For Fig. 4A, N = 85000 and $\alpha$ = 1.15. In above scheme, the random nature of the lattice (for our choice of randomness, the mean coordination number is 6 with standard deviation 1) biases the average aging rate $\rho$ so that a homogeneous population ($\sigma$=0) will have $\rho$ = 1.015. When a hexagonal lattice with no randomness is used, $\rho$=1. A 0.015 offset is added to the theoretical curves to account for this. For Fig.7C (top) N = 200000 and $\alpha$ = 1.25. For each figure, the four curves shown are fitted simultaneously using the same two fitting parameters.

\section*{Acknowledgments}
The author thanks Boris Shraiman, Adel Dayarian, and Sid Goyal for helpful discussions and critical reading of the manuscript, and Arnie Levine for introduction to the subject. This research was supported in part by the National Science Foundation under Grant No. NSF PHY11-25915.






\end{document}